\documentstyle[twocolumn,prl,aps]{revtex}
 
\begin{document}
\wideabs{ 
\title
{Rigorous approach to the problem of ultraviolet divergencies in 
dilute Bose gases}
\author{Maxim Olshanii$^{1}$\cite{e-mail1} and Ludovic Pricoupenko$^{2}$\cite{e-mail2}}
\address
{$^{1}$Department of Physics \& Astronomy,
University of Southern California,
Los Angeles, CA 90089-0484, USA
\\$^{2}$Laboratoire de Physique Th\'{e}orique des Liquides, Universit\'{e} Pierre et Marie Curie,
case 121, 4 place Jussieu, 75252 Paris Cedex 05, France. URA 765 of CNRS}
\date{\today}

\maketitle
\begin{abstract}
In this Letter we consider a system of 
$N$ pairwise finite-range interacting atoms and prove 
rigorously that 
in the zero-range interaction limit all the eigenstates and eigenenergies
of the Hamiltonian 
converge to those corresponding to  
$N$ 
atoms interacting via the Fermi-Huang regularized pseudo-potential. 
Next, we show  that the latter eigensystem (if treated exactly) is {\it invariant} under 
a nontrivial transformation of the interaction potential.
Finally, we realize that most of the approximate schemes of  
many-body physics do not exhibit this invariance:
we use this property to  
resolve all inconsistencies of the Hartree-Fock-Bogoliubov variational
formalism known so far.
\end{abstract}
%
}
	The realization of Bose Einstein Condensate \cite{BEC} has
brought an enormous interest in developing new theoretical approaches 
and refining the existing ones. The mean-field formalism with contact
interactions has been shown to provide a powerful tool for analyzing 
the properties of trapped Bose gases \cite{Stringari}. Unfortunately, 
the most general variational mean field approach: the  
Hartree-Fock-Bogoliubov approximation (HFB), is not yet quite
satisfactory if used with contact interactions: it exhibits UV-divergencies, 
inconsistencies with the Hugenholtz-Pines theorem \cite{Griffin}, many-body 
$T$-matrix calculations \cite{Stoof}, and even
with the very existence of atomic condensates themselves
\cite{toy-model}. Several heuristic modifications of the theory were suggested
\cite{Griffin,Stoof,Burnett,Giorgini}, showing a good agreement 
with the experimental data \cite{JILA-MIT_excitations}.
In this Letter, we propose a novel,
quite straightforward way to cure these inconsistencies.

	First, we prove rigorously that the regularized Fermi-Huang 
pseudo-potential
\cite{Huang} is not just an
ansatz, but provides the exact zero-range limit of the many-body 
observables 
along with a cancellation of all the UV-divergencies. 
Second, we introduce a new family of pseudo-potentials parametrized by a free 
parameter $\Lambda$ (so-called $\Lambda$-potentials): no {\it exact} (after the zero-range 
approximation has been made) observable depends on it, while some {\it approximate} 
treatments differ for different $\Lambda$. 
The above conclusions are  general and they do not rely on any particular approximation.
Finally, as an application of
this new potential, we find a particular value for $\Lambda$ such that HFB 
equations are entirely free of all inconsistencies known so far. 

An explicit expression for the $\Lambda$-potential  reads 
\begin{equation}
\hat{V}^\Lambda(\vec{r}) = g_\Lambda \delta(\vec{r}) 
\left[ \partial_r + \Lambda \right] (r \,\cdot\,)
\quad \mbox{with} \quad g_\Lambda = \frac{g_0}{1-a\Lambda}\,,
\label{eq:vpseudo}
\end{equation}
where $\vec{r}=\vec{r}_1-\vec{r}_2$ is the relative coordinate of two
atoms, $a$ is the $s$-wave scattering length, $g_0 = 2\pi\hbar^2
a/\tilde{\mu}$ is the usual effective coupling constant and 
$\tilde{\mu} = m/2$ is the reduced mass. When $\Lambda=0$, $\hat{V}^{\Lambda=0}$
coincides with the Fermi-Huang pseudo-potential
. For a $1/r$-divergent wave-function $\displaystyle \psi(\vec{r}) =
\tilde{\alpha}/r + \eta(\vec{r})$\cite{regular}, 
the action of the $\Lambda$-potential is
\begin{eqnarray}
\hat{V}^\Lambda \psi(\vec{r}) 
&=& g_{\Lambda} \delta(\vec{r}) \left\lbrack \eta(\vec{0}) + \Lambda\tilde{\alpha} \right\rbrack
\label{eq:pp_action}
\end{eqnarray}
For a low-energy two-body body scattering process, the eigenstates
of the $\Lambda$-potential coincide with  
the asymptotic form of the $s$-wave eigenstates of any other
interaction potential of a scattering length $a$.  
However, 
for energies of the order $\displaystyle \frac{\hbar^2}{ma^2}$ or higher,
the actual
finite size structure 
of the potentials comes into play,
and the range of the applicability
of the zero-range approximation  reduces to
the particular case of a zero-energy resonance.    

	Now, we consider a system of $N$ atoms of mass $m$ and coordinates 
$\left\{ {\bf r} \right\} \equiv\left\{ \vec{r}_1,\vec{r}_2,\ldots,\vec{r}_N \right\}$ 
interacting via $\hat{V}^\Lambda$. The Hamiltonian reads
\begin{equation} 
\hat{\cal H}^{p.p.} =  \sum_{i=1}^N \frac{p_i^2}{2 m} + \sum_{\alpha=1}^{\alpha_{max}} \hat{V}^\Lambda(r_\alpha)
\label{PP_H} \quad ,
\end{equation}   
where $\alpha = 1, \, \ldots, \, \alpha_{max} \equiv N(N-1)/2$ labels ordered pairs of atoms
$(i_\alpha, \, j_\alpha)$,  and $\vec{r}_\alpha = \vec{r}_{i_\alpha}-\vec{r}_{j_\alpha}$
is the relative position of the members of the $\alpha$-th pair.
As a direct consequence of Eq.(\ref{eq:pp_action}), 
any eigenstate $\Psi(\left\{ {\bf r} \right\})$ of
(\ref{PP_H}) is a solution of an interaction-free Schr\"{o}dinger 
equation subject to the following contact conditions for all pair $\alpha$:
\begin{equation}
\lim_{r_{\alpha}\to 0}
\frac{\partial}{\partial r_{\alpha}}\Big|_
{
\left\{  
\mbox{\boldmath${\cal R}$}_{\alpha}
\right\} 
}
\ln\lbrack r_{\alpha} \Psi \rbrack = -\, \frac{1}{a}
\label{eq:contact} \quad ,
\end{equation}
where $\left\{ \mbox{\boldmath${\cal R}$}_\alpha \right\}
\equiv \left\{  \vec{R}_{\alpha},\, \{ \vec{r}_i|\, i \neq i_\alpha,j_\alpha \} \right\}$
is a set composed of the coordinate of  the center of mass 
$\vec{R}_{\alpha}=(\vec{r}_{i_\alpha}+\vec{r}_{j_\alpha})/2$ of the $\alpha$-th pair and 
all other coordinates not belonging to this pair.
Indeed, it is easy to show that the $\delta$-singularities
in the action of the Hamiltonian (\ref{PP_H})
on a many-body state $\Psi(\{\vec{r}\})$ cancel each other 
if and only if
$\Psi$ satisfies the contact conditions  (\ref{eq:contact}).
Notice that these contact conditions  
{\it do not depend on $\Lambda$}, and thus no {\it exact} (after the zero-range approximation 
has been made) eigenstate does. 

Now, we are going to show how regularized pseudo-potentials arise 
in the limit of zero-range interactions. For this purpose,
we consider $N$ particles  of mass $m$, interacting via a potential which 
belongs to a one-parameter family of square-well potentials:
$v^R(r) = - v_0^{R}\, \Theta(R-r)$. The depth of the potential $v_0^{R}>0$
is chosen in such a way that the scattering length $a$ is the same 
for all members of the family, and each of them supports 
the same number of $s$-wave bound states,
either one or none depending on 
the sign of the scattering length \cite{implicit_v0}. The Hamiltonian of the system reads  
\begin{eqnarray}
\hat{\cal H}^{R} = \sum_{i=1}^{N} \frac{p_i^2}{2 m}
 + \sum_{\alpha=1}^{\alpha_{max}} v^{R}(r_\alpha) 
\label{total_H}
\quad .
\end{eqnarray}
%
We wish  to prove the following 
\\
\noindent
{\it STATEMENT.- In the limit of infinitely small potential range, the    
Green's function of the finite-range-interaction Hamiltonian (\ref{total_H})
converges to the Green's function of the pseudo-potential Hamiltonian (\ref{PP_H}):}
\begin{equation}
\lim_{R \to 0} (E - \hat{\cal H}^R)^{-1}=(E - \hat{\cal
H}^{p.p.})^{-1} \quad .
\label{statement}
\end{equation}   
{\it Proof.} As we saw above (see (\ref{eq:contact})), all the $\Lambda$-potentials lead to the same 
eigenstates (and thus the same Green's function): hence,
without loss of generality, we can limit
our proof to the case of the Fermi-Huang pseudo-potential $\hat{V}^{0}$. 

Let us define two operator-valued functions:
\begin{equation}
\hat{\cal G}_{\hat{a}}^E = (E + i\epsilon - \hat{a})^{-1}
\quad , \quad
\hat{\cal T}_{\hat{a}, \hat{b}}^E = (1 - \hat{b}\hat{\cal G}_{\hat{a}}^E)^{-1} \hat{b}
\quad .
\end{equation}
The former is the retarded Green's function at energy $E$ for a Hamiltonian $\hat{a}$.
The latter is the $T$-matrix of a perturbation $\hat{b}$ in presence of the
background Hamiltonian $\hat{a}$. Two relations will be heavily used in what follows. First is
the Lippman-Schwinger relation between the Green's function of the ``full Hamiltonian''
$\hat{a} + \hat{b}$ and the one of the background:
\begin{equation}
\hat{\cal G}_{\hat{a}+\hat{b}}^E = \hat{\cal G}_{\hat{a}}^E +
\hat{\cal G}_{\hat{a}}^E \hat{\cal T}_{\hat{a}, \hat{b}}^E \hat{\cal
G}_{\hat{a}}^E \quad .
\label{L-SCH}
\end{equation}
The second is the Lupu-Sax formula \cite{Lupu-Sax} relating the $T$-matrices of the
same perturbation but in two different background Hamiltonians $\hat{a}_1$ and $\hat{a}_2$:
\begin{equation}
\hat{\cal T}_{\hat{a}_2, \hat{b}}^E =\lbrack
1 - \hat{\cal T}_{\hat{a}_1, \hat{b}}^E
    (\hat{\cal G}_{\hat{a}_2}^E - \hat{\cal G}_{\hat{a}_1}^E ) \rbrack^{-1}
    \hat{\cal T}_{\hat{a}_1, \hat{b}}^E \quad .
\label{Adam}
\end{equation}
Introduce also a family of ``reduced Hamiltonians'' $\hat{\cal
H}^R_{\{\alpha\}}$ and  a family of reference Hamiltonians $\hat{h}^{E}_\alpha$ 
\begin{eqnarray} 
\hat{\cal H}^R_{\{\alpha\}} =  \sum_{i=1}^N \frac{p_i^2}{2 m} + \sum_{\beta=1}^\alpha
v^R(r_\beta) \quad 
\mbox{and} \quad \hat{h}^{E}_\alpha = \frac{p_\alpha^2}{2\tilde{\mu}} + E , 
\label{reduced_H}\label{ref_Hamiltonian}
\end{eqnarray}   
where $\vec{p}_\alpha =(\vec{p}_{i_\alpha} - \vec{p}_{j_\alpha})/2$
is the relative momentum for the $\alpha$-th pair. Each reference Hamiltonian 
is just a sum of the relative kinetic energy for the 
corresponding pair and the energy $E$ at which the 
Green's functions (\ref{statement}) are compared.
The Green's function of the $\alpha$-th reference Hamiltonian
is proportional to the zero-energy Green's function for the 
relative motion of two particles:
\begin{equation}
\langle \left\{ {\bf r} \right\} |
     \hat{\cal G}^E_{\hat{h}^{E}_\alpha}
  | \left\{ {\bf r}^\prime \right\}  \rangle
= - \frac{\tilde{\mu}}{2\pi \hbar^2 |\vec{r}_\alpha-\vec{r}^{\ \prime}_\alpha|} \,
  \delta(\left\{\mbox{\boldmath${\cal R}$}_\alpha -
          \mbox{\boldmath${\cal R}$}^\prime_\alpha\right\})
\,.
\label{2B-Green}  
\end{equation}
In turn the $T$-matrix of the interaction potential $\hat{v}^R_\alpha \equiv v^R(r_\alpha)$
in presence of $\hat{h}^{E}_\alpha$ can be expressed 
through the {\it zero-energy two-body $T$-matrix} of it:
\begin{equation}
\langle \left\{ {\bf r} \right\} |
    \hat{\cal T}_{\hat{h}^{E}_\alpha,\hat{v}^R_\alpha}^E
| \left\{ {\bf r}^\prime \right\} \rangle = 
g_0 D^{R}(\vec{r}_\alpha,\vec{r}_\alpha^{\ \prime}) \,
  \delta(\left\{\mbox{\boldmath${\cal R}$}_\alpha- 
           \mbox{\boldmath${\cal R}$}^\prime_\alpha \right\}) .
\label{ref_T-matrix}
\end{equation}
The kernel $D^R$ is zero when $r_\alpha>R$ or $r_\alpha'>R$ and 
is normalized to unity as $ \int d^3\vec{r} d^3\vec{r}^{\,\prime} \, D^R(\vec{r},\vec{r}^{\,\prime}) = 1 $. 
An explicit expression for it can be found in 
Ref.\cite{kernel}. In the limit of zero-range interaction, the kernel 
obviously converges to a product of delta-functions, and hence the $T$-matrix converges to:
\begin{eqnarray}
\hat{\cal T}_{\hat{h}^{E}_\alpha,\hat{v}^R_\alpha}^E
\stackrel{R\to 0}{\longrightarrow} g_0 \delta(\vec{r}_\alpha)
\quad .
\label{2B-T}
\end{eqnarray} 
Notice that by construction of the reference Hamiltonian $\hat{h}^{E}_\alpha$ 
neither the Green's function $\hat{\cal G}^E_{\hat{h}^{E}_\alpha}$ nor the $T$-matrix  
$\hat{\cal T}_{\hat{h}^{E}_\alpha,\hat{v}^R_\alpha}^E$
depend on energy $E$.

Using relations (\ref{L-SCH}, \ref{Adam}), the full many-body 
Green's function of the system can be rigorously expressed through 
the zero-energy two-body $T$-matrices (\ref{ref_T-matrix}) of the interaction potential 
$v^R(r)$. Removing from the Hamiltonian (\ref{total_H}) one pair interaction after another, 
we obtain the following chain relation:
\begin{eqnarray}
&&
(E - \hat{\cal H}^{R})^{-1} \equiv
\hat{\cal G}^{E}_{\hat{\cal H}^{R}} \equiv \hat{\cal G}^{E}_{\hat{\cal H}^{R}_{\{\alpha_{max}\}}}
\nonumber
\\
&
\uparrow
&
\,\,\cdots
\label{Down-Up}
\\
&&
\hat{\cal G}^E_{\hat{\cal H}^R_{\{\alpha\}}} =
      \hat{\cal G}^E_{\hat{\cal H}^R_{\{\alpha-1\}}} 
\nonumber
\\
&&
\hspace{1cm}
      +
      \hat{\cal G}^E_{\hat{\cal H}^R_{\{\alpha-1\}}}
      \lbrack 1 - \hat{\cal T}^E_{\hat{h}^{E}_{\alpha},\hat{v}^R_{\alpha}}
      (\hat{\cal G}^E_{\hat{\cal H}^R_{\{\alpha-1\}}}
      - \hat{\cal G}^E_{\hat{h}^{E}_{\alpha}}) \rbrack^{-1}  
\nonumber
\\
&&
\hspace{5cm}
      \times
      \hat{\cal T}^E_{\hat{h}^{E}_{\alpha},\hat{v}^R_{\alpha}} 
      \hat{\cal G}^E_{\hat{\cal H}^R_{\{\alpha-1\}}} 
\nonumber
\\
&
\uparrow
&
\,\,\cdots
\nonumber
\\ 
&&
\hat{\cal G}^{E}_{\hat{\cal H}_{\{0\}}} \equiv
(E - \sum_{i=1}^{N} \frac{p_{i}^2}{2 m})^{-1} 
\quad .
\nonumber
\end{eqnarray}   
This relation is the cornerstone of the proof.

Imagine now that the full Green's function  $\hat{\cal G}^E_{\hat{\cal H}^R}$
acts on a state $|\Psi\rangle$ whose wave function $\Psi(\left\{ {\bf r} \right\})$ is 
regular everywhere. Computation of the result of this action 
$\langle \left\{ {\bf r} \right\} | \hat{\cal G}^E_{\hat{\cal H}^R} | \Psi\rangle$ 
involves expressions of a form 
$
\hat{\cal T}^E_{\hat{h}^{E}_\alpha,\hat{v}^R_\alpha}
  (\hat{\cal G}^E_{\hat{\cal H}^R_{\{\alpha-1\}}}
  -\hat{\cal G}^E_{\hat{h}^{E}_\alpha})
\hat{\cal T}^E_{\hat{h}^{E}_\alpha,\hat{v}^R_\alpha}
\Xi
$
where $\Xi(\left\{  {\bf r} \right\})$ is a regular function,
and in the limit $R \to 0$, leads to the expressions of the following type:
\begin{eqnarray}
g_0 \delta(\vec{r}_\alpha) \left\lbrack  \Gamma(\vec{r}_\alpha) -
\frac{\tilde{\alpha}}{r_\alpha} \right\rbrack
\label{dum_1} \quad .
\end{eqnarray}
(In Eq.(\ref{dum_1}), both $\Gamma(\vec{r}_\alpha)$ and $\tilde{\alpha}$ depend also on 
$\left\{\mbox{\boldmath${\cal R}$}_{\alpha} \right\}$.) $\Gamma$ is given by 
\begin{eqnarray}
\Gamma(\vec{r}_\alpha) = \int d^{3N}\left\{ {\bf r}^\prime \right\} 
\langle \left\{ {\bf r} \right\} | \hat{\cal G}^E_{\hat{\cal H}^{R \to 0}_{\{\alpha-1\}}} | 
\left\{ {\bf r}^\prime \right\} \rangle g_0 \delta(\vec{r}^{\ \prime}_\alpha) \Xi(\left\{ {\bf
r}^\prime \right\}) ,
\nonumber
\end{eqnarray}
and 
$
\displaystyle \tilde{\alpha} = 
- g_{0}\, (\tilde{\mu}/2\pi\hbar^2)\, \Xi(\left\{  {\bf r} \right\}) \Big|_{\vec{r}_{\alpha}=0} 
$. Now using the definition of the Green's function $\hat{\cal
G}^E_{\hat{\cal H}^R_{\{ \alpha-1\}}}$,
we find that $\Gamma(\vec{r}_\alpha)$ has an UV singularity of 
form $\tilde{\alpha}/r_\alpha$ (the same as the second term in the expression (\ref{dum_1}) has)
(see \cite{proof_2}).
This leads to 
\begin{eqnarray} 
g_0 \delta(\vec{r}_\alpha) \left\lbrack 
 \Gamma(\vec{r}_\alpha) - \frac{\tilde{\alpha}}{r_\alpha} \right\rbrack
=\hat{V}^0_\alpha \Gamma(\vec{r}_\alpha) \quad ,
\end{eqnarray}   
{\it i.e.} the expression (\ref{dum_1}) involves the Fermi-Huang
pseudo-potential 
$\hat{V}^0_\alpha \equiv \hat{V}^{0}(r_{\alpha})$ ({\it c.f.} (\ref{eq:pp_action}) at $\Lambda = 0$). 
This justifies the following limit:
\begin{equation}
\hat{\cal T}^E_{\hat{h}^{E}_\alpha,\hat{v}^R_\alpha} 
             (\hat{\cal G}^E_{\hat{\cal H}^R_{\{\alpha-1\}}}-
             \hat{\cal G}^E_{\hat{h}^{E}_{\alpha}})
\stackrel{R\to 0}{\to}
  \hat{V}^0_\alpha  \hat{\cal G}^E_{\hat{\cal H}^{R\to 0}_{\{\alpha-1\}}} \quad .
\label{dum_2}
\end{equation}   
Inserting the above substitution at every level of the 
chain procedure (\ref{Down-Up}) and collecting all the terms  
\cite{end-terms}
one finally arrives at 
$\displaystyle \lim_{R\to 0} \hat{\cal G}^{E}_{\hat{\cal H}^R} 
= \hat{\cal G}^{E}_{\hat{\cal H}^{p.p.}}$, {\it Q.E.D.}
.

Notice that the relation (\ref{dum_2}) clearly shows that the role of the 
regularizing operator in the pseudo-potential expression
(\ref{eq:vpseudo}) is to subtract the free propagators $\hat{\cal G}^E_{\hat{h}^{E}_{\alpha}}$
already taken into account by the two-body $T$-matrix $\hat{\cal
T}^E_{\hat{h}^{E}_\alpha,\hat{v}^R_\alpha}$. As a result, UV divergencies
disappear at each level of the chain recursion (\ref{Down-Up}).

As an application of the $\Lambda$-potential, we consider now 
the HFB theory for N bosons interacting via $V^\Lambda$ with $a>0$, in a box of a size $L$.  
As we will see, the $\Lambda$-freedom in choosing the effective Hamiltonian (\ref{PP_H})
offers the following advantages: 
(a) Unlike for the conventional HFB formalism ($\Lambda = 0$), there exists a
range of $\Lambda$ such that the {\it atomic} condensate constitutes 
the minimum of the HFB functional in the low density regime \cite{Lambda<star};
(b) For a particular value, $\Lambda = \Lambda^{\star}$, HFB equations are
consistent with the results of the ladder approximation for the many-body $T$-matrix \cite{Stoof}
and the Hugenholtz-Pines theorem is
satisfied
;
(c) In the vicinity of  $\Lambda^{\star}$ the ground state energy of the system 
is consistent with Bogoliubov's predictions.  

The HFB approximation is twofold. First, it breaks the $U(1)$ symmetry:
the atomic field $\hat{\psi}$ is split into a classical field $\Phi$ and a 
quantum fluctuation $\hat{\phi} = \hat{\psi} - \Phi$. Second, the exact density operator
is replaced by a Gaussian variational ansatz: $\hat{D} \equiv \exp(-\hat{K}/k_{\rm
B}T)/Z$, where $Z$ is the partition function and the quadratic
variational Hamiltonian is
\begin{eqnarray} 
\hat{K}[h,\,\Delta,\,\Phi] 
&=& \frac{1}{2} \int\!\int\!d^3\vec{r_1} d^3\vec{r_2} \ \big[ \hat{\phi}^{\dagger}(\vec{r}_1)
h(\vec{r_1},\vec{r_2}) \hat{\phi}(\vec{r}_2) \nonumber \\
& & + \ \hat{\phi}^\dagger(\vec{r}_1) \Delta(\vec{r}_1,\vec{r}_2) \hat{\phi}^\dagger(\vec{r}_2) +
{\rm h.c.} \big] \quad . \label{eq:ansatz}
\end{eqnarray}
For what follows, we introduce the coordinates
$\vec{R}=(\vec{r}_1+\vec{r}_2)/2$ and
$\vec{r}=\vec{r}_1-\vec{r}_2$. 
The second-quantized form of the full Hamiltonian (\ref{PP_H})   
reads
\begin{eqnarray}
\hat{\cal H}^{\Lambda}
 &=&
    \int\! d^3\vec{R} \,
 \left\{
           \hat{\psi}^{\dagger}
           ( - \frac{\hbar^2}{2m} \Delta)
           \hat{\psi}
     +
      \frac{g_{\Lambda}}{2}\,
      \hat{\psi}^{\dagger}
      \hat{\psi}^{\dagger}
      \stackrel
      {  
       \stackrel
        {
        \Lambda
        }
        {
        \unitlength=1em\line(0,-1){.3}\line(+1,0){1.1}\line(0,-1){.3}
        }
      }  
      {  
      \hat{\psi}
      \, 
      \hat{\psi}
      }  
 \right\} ,
\label{bose_Hamiltonian}
\end{eqnarray}
where
\begin{eqnarray} 
      F( 
      \stackrel  
      {  
       \stackrel 
        {
        \Lambda  
        }
        {
        \unitlength=1em\line(0,-1){.3}\line(+1,0){1.7}\line(0,-1){.3}
        }
      }  
      { 
      \vec{R},\,
      \vec{R}
      } 
      )
 =      
      \lim_{r \to 0}
      \left\{
      \partial_r + \Lambda
      \right\} 
      \left\lbrack
      r\, F(\vec{R}+\vec{r}/2,\, \vec{R}-\vec{r}/2)
      \right\rbrack
 . \label{regularization_operator}
\end{eqnarray}   
is a shortened notation for the action of the regularizing operator
(see (\ref{eq:vpseudo})).
Using Wick's theorem, we obtain an
approximate grand canonical potential $J\equiv E^\Lambda-\mu N-TS$, where 
$E^\Lambda=\mbox{Tr}[\hat{\cal H}^{\Lambda} \hat{D}]$ is the energy, 
$N$ is the number of particles, and 
$S = -k_{\rm B}\mbox{Tr}[\ln(\hat{D}) \hat{D}]$ is the
entropy. Minimization of $J$ with respect to the three
variational fields $h,\Delta$ and $\Phi$ 
leads to the following implicit equations for these fields:  
\begin{eqnarray} 
h(\vec{r_1},\vec{r_2}) &=& -\frac{\hbar^2}{2m}(\vec{\nabla}^2\delta)(\vec{r})
+ \left[ \hbar\Sigma_{11}^{\Lambda} - \mu \right] \delta(\vec{r})  \nonumber\\
\Delta(\vec{r}_1,\vec{r}_2) &=& \hbar\Sigma_{12}^{\Lambda} \delta(\vec{r}) \label{eq:HFB}\\
-\frac{\hbar^2}{2m} \Delta \Phi &+& \left[ g_{\Lambda}
(2 \tilde n + |\Phi|^2) - \mu \right] \Phi + g_{\Lambda} \tilde \kappa_\Lambda \Phi^* = 0 
\,,
\nonumber
\end{eqnarray}
where
$\hbar \Sigma_{11}^{\Lambda} = 2 n g_\Lambda $ 
and 
$\hbar \Sigma_{12}^{\Lambda} = g_\Lambda (\Phi^2 + \tilde{\kappa}_\Lambda)$
are the self-energies, $\tilde{n} = \mbox{Tr}[\hat{\phi}^\dagger(\vec{R})\hat{\phi}(\vec{R})\hat{D}]$
is the non-condensed density, $n = |\Phi|^2 + \tilde{n}$ is the total
density, and
$
\tilde{\kappa}_\Lambda = 
      \tilde{\kappa}
      (
      \stackrel  
      {  
       \stackrel 
        {
        \Lambda  
        }
        {
        \unitlength=1em\line(0,-1){.3}\line(+1,0){1.7}\line(0,-1){.3}
        }
      }  
      {  
      \vec{R},\, 
      \vec{R}
      }  
      )
$
results from the action of the regularizing operator 
(\ref{regularization_operator}) on the anomalous density 
$\tilde{\kappa}(\vec{r}_1,\, \vec{r}_2) = \mbox{Tr}[\hat{\phi}(\vec{r}_1)\hat{\phi}(\vec{r}_2)\hat{D}]$. 

The diagonalization of the variational Hamiltonian $\hat{K}$ leads to the following 
quasi-particle spectrum: 
\begin{eqnarray}
\hbar\omega_k = \left( \frac{\hbar^2 k^2}{2m} + 2 g_\Lambda \Phi^2
\right)^{\frac{1}{2}} \left( \frac{\hbar^2 k^2}{2m} - 2 g_\Lambda \tilde{\kappa}_\Lambda \right)^{\frac{1}{2}}
\quad.
\label{eq:spectrum}
\end{eqnarray}
Eqs.(\ref{eq:HFB},\ref{eq:spectrum}) clearly show that HFB is 
$\Lambda$-dependent. As it has been shown 
in Ref.\cite{Stoof}, this approach is only able to provide a Born approximation
for the diagonal self-energy $\hbar \Sigma_{11}$, 
hence its explicit $\Lambda$-dependence. However, this is not the
case for  $\hbar \Sigma_{12}$; indeed, the total pairing field reproduces 
the contact conditions (\ref{eq:contact}) of a two-body wave function
\begin{equation}
\langle \hat{\psi}(\vec{r}_1) \hat{\psi}(\vec{r}_2) \rangle 
= (\Phi^2 + \tilde{\kappa}_{0}) \left( 1-\frac{a}{r} \right) + {\cal O}(r)
\quad,
\end{equation} 
and as a result $\hbar \Sigma_{12}^\Lambda=g_0(\Phi^2+\tilde{\kappa}_0)$ for all $\Lambda$. 

Requiring that all the eigen-energies (\ref{eq:spectrum}) are
real, we find that for zero temperature and densities below a value 
of $\displaystyle n_{\rm crit} = \frac{\pi}{192a^3}$
the existence of an atomic condensate ($\Phi \neq 0$)
implies the following constraint on $\Lambda$:
\begin{equation}
\Lambda^{\star}	a \leq \Lambda a<1 \quad \mbox{with} \quad \Lambda^{\star} a= \frac{\tilde
\kappa_0}{\Phi^2 + \tilde{\kappa_0}} \quad .
\label{Lambda_star}
\end{equation}
At the lower limit $\Lambda = \Lambda^{\star}$, the $\Lambda$-regularized anomalous density disappears,
and the theory
become {\it fully consistent} with 
the results of the many-body 
$T$-matrix calculations in the ladder diagrams approximation
\cite{Stoof} 
%
\begin{eqnarray} 
&&
\tilde{\kappa}_{\Lambda^{\star}} =0 \quad ; \quad
\hbar \Sigma_{11}^{\Lambda^{\star}} = 2 n g_{\Lambda^{\star}} \quad;\quad 
\hbar \Sigma_{12}^{\Lambda^{\star}} = g_{\Lambda^{\star}} \Phi^2
\nonumber
\\
&&
g_{\Lambda^{\star}} = g_{0}[1+\frac{\tilde{\kappa}_{0}}{\Phi^2}] 
= T^{\rm MB}(\vec{0},\,\vec{0},\,\vec{0};\,0)
\quad ,
%
\end{eqnarray}   
yielding a gapless spectrum \cite{G2,gap}. 

Consider now the zero-temperature low-density limit of our equations.
Assuming $\Lambda a$ to be of the order of $\sqrt{na^3}$
and neglecting all the terms of order $na^3$ or higher, the energy $E^\Lambda$ is
{\it independent} of $\Lambda$ and coincides with the well known Bogoliubov's result
\begin{equation}
E^{\Lambda}= \frac{g_{0}}{2}nN \left(1 + \frac{128}{15\sqrt{\pi}} \sqrt{na^3} + \ldots\right) \quad .
\label{E_Bogoliubov}
\end{equation}
The $\Lambda$-potential based variational HFB model is therefore consistent with the
perturbative Bogoliubov's approach. As the density increases 
the parameter $\Lambda^{\star}$ increases as well, 
and at a critical density 
$\displaystyle n_{\rm crit} = \frac{\pi}{192a^3}$  
we find $\Lambda^{\star} a = 1$: the energy diverges and the mean field
treatment breaks down. 

Note in conclusion that the $\Lambda$-invariance described in our Letter
holds even if the constant $\Lambda$ is replaced by an arbitrary field
$\Lambda(\vec{R})$. The generalization of our HFB theory to the case 
of the trapped gases is thus straightforward:
One has simply to fix $\Lambda$ as $\Lambda(\vec{R})=\Lambda^{\star}(\tilde{\kappa}_0(\vec{R}), \Phi^2(\vec{R}))$
according to (\ref{Lambda_star}) at every point  $\vec{R}$ of the trap.

As a extension of this work, we mention that using a procedure 
similar to the 3D case, it is possible to obtain the 
low-dimensional analogs of the $\Lambda$-potential:  
\begin{eqnarray}
&&
V_{\rm 2D}^{\Lambda}(\vec{\rho}) =
-
\frac{\pi\hbar^2}{\tilde{\mu}} \,
\frac{1}{\log(q\Lambda R)} \,
\delta(\vec{\rho})\,
\left\{
1
-
\log(q\Lambda \rho)
\rho\frac{\partial}{\partial \rho}
\right\}
\nonumber
\\
&&
V_{\rm 1D}^{\Lambda}(z) =-\frac{\hbar^2}{\tilde{\mu}} \,
\frac{\Lambda}{\Lambda a_{\rm 1D} - 1 } \,
\delta(z)\,
\left\{
1
+
\frac{1}{2\Lambda} \,
\frac{\partial}{\partial z}
(
\Big|_{0+} \!\!\!\!- \Big|_{0-}
)
\right\} ,
\nonumber
\end{eqnarray}
where $q = e^{C}/2$, $C$ is the Euler's constant, $R$ 
is the 2D effective hard disk radius and $a_{\rm 1D}$ is 
the 1D scattering length \cite{2D-T,2D-1D}.

{\bf Acknowledgments}. Authors are grateful to A.~Lupu-Sax, 
R.~Shakeshaft, R.~Thompson, and especially to 
Y.~Castin for
enlightening discussions on the subject. This work was supported by 
the NSF grant {\it PHY-0070333}.


\begin{thebibliography}{99}
\bibitem{regular} $\eta(\vec{r})$ is a regular function and 
$
	\eta(\vec{0})= \lim_{r\to 0} \left[ \psi(\vec{r}) -
\frac{\tilde{\alpha}}{r} \right] \quad ,
$
is the regular part of the wavefunction.
\bibitem{implicit_v0} $v_0$ is the smallest positive solution of 
the implicit equation $a = R(1 - \tan(k_0R)/(k_0R))$, where
$k_0 = \sqrt{m v_0}/\hbar$.
\bibitem{end-terms} This calculation also involves 
expressions of a form
$\hat{\cal T}^E_{\hat{h}^{E}_\alpha,\hat{v}^R_\alpha}\Xi$
where the limit 
$
\hat{\cal T}^E_{\hat{h}^{E}_\alpha,\hat{v}^R_\alpha}\Xi 
\stackrel{R\to 0}{\to}
\hat{V}^0_\alpha\Xi
$ 
is also well justified.
\end{thebibliography}
\end{document}